\documentclass{article}

\usepackage{arxiv}

\usepackage[utf8]{inputenc} 
\usepackage[T1]{fontenc}    
\usepackage{hyperref}       
\usepackage{url}            
\usepackage{booktabs}       
\usepackage{amsfonts}       
\usepackage{nicefrac}       
\usepackage{microtype} 
\usepackage{doi}

\usepackage[]{graphicx}       
\usepackage{amssymb,amsmath, amsthm}        
\usepackage[table,xcdraw]{xcolor} 
\usepackage{multirow} 
\usepackage{textcomp, gensymb}
\usepackage[utf8]{inputenc}

\title{A Bayesian spatio-temporal study of association between meteorological factors and the spread of COVID-19}

\author{ \href{https://orcid.org/0009-0007-7424-7515}{\hspace{1mm}Jamie Mullineaux} \\
	Department of Statistical Science\\
	    University College London\\
	London, United Kingdom \\
	\texttt{james.mullineaux.21@ucl.ac.uk} \\
	\AND
	\href{https://orcid.org/0000-0001-6420-6567}
     {\hspace{1mm}Baptiste Leurent} \\
	Department of Statistical Science\\
	    University College London\\
	London, United Kingdom \\
	\texttt{baptiste.leurent@ucl.ac.uk}
	\And
	\href{https://orcid.org/0000-0001-7846-9763}{\hspace{1mm}Takoua Jendoubi} \\
	Department of Statistical Science\\
	    University College London\\
	London, United Kingdom \\
	\texttt{t.jendoubi@ucl.ac.uk} \\
}

\begin{document}
\maketitle

\begin{abstract}
The spread of COVID-19 has brought challenges to health, social and economic systems around the world. With little to no prior immunity in the global population, transmission has been driven primarily by human interaction. However, as with common respiratory illnesses such as influenza some authors have suggested COVID-19 may become seasonal as immunity grows. Despite this, the effects of meteorological conditions on the spread of COVID-19 are poorly understood. Previous studies have produced contrasting results, due in part to limited and inconsistent study designs. This study investigates the effects of meteorological conditions on COVID-19 infections in England using a Bayesian conditional auto-regressive spatio-temporal model. Our data consists of daily case counts from local authorities in England during the first lockdown from March – May 2020. During this period, legal restrictions limiting human interaction remained consistent, minimising the impact of changes in human interaction. We introduce a lag from weather conditions to daily cases to accommodate an incubation period and delays in obtaining test results. By modelling spatio-temporal random effects we account for the nature of a human transmissible virus, allowing the model to isolate meteorological effects. Our analysis considers cases across England's 312 local authorities for a 55-day period. We find relative humidity is negatively associated with COVID-19 cases, with a 1\% increase in relative humidity corresponding to a reduction in relative risk of 0.2\% (95\% highest posterior density [HPD]: 0.1\% - 0.3\%). However, we find no evidence for temperature, wind speed, precipitation or solar radiation being associated with COVID-19 spread. The inclusion of weekdays highlights systematic under reporting of cases on weekends with between  27.2\% - 43.7\% fewer cases reported on Saturdays and 26.3\% - 44.8\% fewer cases on Sundays respectively (based on 95\% HPDs). By applying a Bayesian conditional auto-regressive model to COVID-19 case data we capture the underlying spatio-temporal trends present in the data. This enables us to isolate the main meteorological effects and make robust claims about the association of weather variables to COVID-19 incidence. Overall, we find no strong association between meteorological factors and COVID-19 transmission.

\end{abstract}

\keywords{COVID-19 \and Spatio-temporal \and CARBayesST \and Bayesian \and Humidity \and Meteorological}

\section{Introduction}
\subsection{Background}
Since January 2020 COVID-19 has spread across the world and changed societies dramatically. As the virus spread many countries took unprecedented measures to control it, including restrictions on public gatherings and movement. This had negative knock-on economic and societal impacts as well as creating mental health issues. A UK government study showed deterioration in public mental health during lockdown periods \cite{mentalhealth}, whilst restrictions led to an increase in the court case backlog \cite{courtbacklog} and demand on the NHS led to a reduction in elective care, substantially increasing the NHS backlog \cite{england2022delivery}. Given the impact both the virus and control measures have had on daily lives, it is crucial for governing, health and economic authorities to understand risk factors associated with COVID-19 so that public policies are designed to balance risks to public health with social and economic impacts.

In response, many studies were conducted to understand key risk factors associated with COVID-19 transmission. Whilst age is known to be a key driver of serious illness other factors such as poverty and ethnicity were also identified as playing an important role in the spread of COVID-19 \cite{mahase2020covid}. These were often linked to increased human interaction, the foremost driver of COVID-19 spread early in the pandemic, due to confounding factors such as living conditions or employment type. However, the effects of meteorological conditions are less well understood. For instance, due to similarities between COVID-19 and the flu many authorities claimed the virus would be less effective in warmer weather \cite{abctrump, bbctwitter, skybolsonaro}. However, these seem ill-judged as the virus has impacted all but the most remote places on the planet. Discrepancies early in the pandemic could be due to factors such as when the disease first arrived in certain countries and the ability of more developed countries to roll out mass testing quicker than less developed countries, many of which are in warmer climates.

Nonetheless, many other respiratory viruses, including influenza and other human coronaviruses (of which COVID-19 is an instance) are susceptible to weather \cite{neher2020potential, kissler2020projecting}. This is due to the virus being less stable in warmer and more humid weather, as well as these conditions limiting the distance the virus can travel, and the susceptibility of the host \cite{lowen}. With COVID-19 likely to become endemic in the longer term, many scientists believe that COVID-19 may display similar behaviour and become seasonal, once immunity has reached a critical point either naturally or via vaccination \cite{neher2020potential, kissler2020projecting, bmj}.

At the time of writing, the World Health Organisation has declared that COVID-19 is no longer a global health emergency, but still classes it as a pandemic \cite{UN}. Whilst life has returned to normal in many countries, this is reflective of the threat that COVID-19 continues to pose. As such it is critical to understand the longer-term behaviour of COVID-19 so that authorities can prepare, for instance by timing booster vaccine programmes for vulnerable sections of the population, or implementing testing infrastructure so that the prevalence and spread of COVID-19 can be tracked.

However, when studying the association of weather with COVID-19 spread many non-meteorological factors that contribute it must be accounted for. In particular, as COVID-19 spreads through human interaction spatial and temporal correlations are present in case data. As such statistical models that capture these spatio-temporal effects are needed to address this question.

To date, previous studies have delivered heterogeneous findings. For instance, a study in Germany found temperature is negatively correlated with the spread of COVID-19, and argued humidity was positively correlated \cite{ganegoda2021interrelationship}. In contrast, a US study found temperature, humidity and precipitation were all negatively correlated with the number of cases \cite{chien2021lagged}. More recently \cite{Ai2022} showed that the effect of meteorological conditions on COVID-19 transmission varies by country, with both low and high ranges of humidity associated with increased transmission in differing climates.

This study highlights the importance of methods used as, unlike many studies which assume monotonic relationships between variables, it allows for non-monotonicity between temperature and case counts. This could be beneficial if, within a normal range of meteorological parameters, there is a sub-range where COVID-19 is most or least transmissible.

One reason suggested for the widespread heterogeneity is the lack of consistency in study designs and analysis methods \cite{kerr2021associations}. Analysis of COVID-19 case data is not straightforward, as methods need to take account of the temporal and geographical dependency between the data points. Numerous approaches are possible, each requiring its own assumptions, opening up the possibility for different conclusions to be drawn from similar data.  In some instances, the dependencies were altogether ignored, resulting in underestimated uncertainty and increased the chance of spurious findings.

In this paper we investigate the relationship between weather conditions and COVID-19 spread. We study the average effect of several variables in order to understand the potential long-term seasonal behaviour of COVID-19. Using a Bayesian spatio-temporal model we capture auto-correlation in case counts. This enables us to separate out the effects of regression parameters and make strong claims about these effects. Recent developments have made such models accessible to researchers, in particular through the use of the \textit{CARBayesST} package in R, introduced by \cite{lee2018spatio}. These models have previously been used in other epidemiological contexts including the study of dengue fever \cite{aswi2021effects}, malaria \cite{rouamba2020bayesian} and human foot and mouth disease \cite{du2018bayesian}.

\newpage 

\section{Data}
\subsection{Study area and period}
The study considers England's $K = 312$ local authority districts (LADs) as of April 2020, including the Isle of Wight and Isles of Scilly. We consider a 61-day period from  $31^{\mathrm{st}}$ March 2020 - $30^{\mathrm{th}}$ May 2020 which has been chosen to reflect the first full lockdown in England when restrictions remained stable. This allows us to assume the nature of social interactions is consistent throughout the study period, not affected by the weather. Therefore we can isolate the natural effects of weather on COVID-19 spread without accounting for indirect effects due to weather affecting social interactions.

\subsection{Daily cases \& geographical region}
Counts of lab-confirmed cases of COVID-19 (daily cases) in England were obtained from Public Health England \cite{pheCases} for each day in the study period. Daily cases were recorded in each lower tier local authority (LTLA) in England and refer to the date when they were confirmed.

We aggregated LTLA daily cases to the LAD level using definitions from the Office for National Statistics (ONS). This aggregation was necessary as LAD shape files for spatial analysis are available for the LADs (but not the LTLAs) through the Open Geography portal \cite{onsGeography}. We used the boundaries that were current as of May 2020. Based on these boundaries, daily cases for the LTLAs of Aylesbury Vale, Chiltern, South Bucks, and Wycombe were aggregated to obtain cases for the LAD of Buckinghamshire. We also merged the LAD boundaries for Hackney and the City of London (Hackney) as well as Cornwall and Isles of Scilly, as daily cases were recorded at this aggregated level. The resulting set of boundaries were used in the analyses to define shared borders between pairs of LADs as well as for presentation of results.

Population data for LADs in England were derived from the ONS mid-2019 annual estimates \cite{onsPopulation}. These were aggregated as above for Hackney with City of London and Cornwall with Isles of Scilly.

\subsection{Meteorological}
 We obtained weather data for the United Kingdom for the whole of 2020 from the European Commission re-analysis database (ERA5) \cite{era5}. This publicly available resource provides an accurate set of hourly data at a 25-km × 25-km grid resolution and includes the following features: solar radiation, temperature, humidity, wind velocity and precipitation. Note here that the spatio-temporal resolutions of the raw weather data do not align with the daily reporting of cases in LADs. ERA5 provides a suite of hourly weather parameters that may affect local COVID-19 transmissions at a 30-km spatial resolution.

We first aggregated the hourly measurements for each grid to the daily level by taking an average from midnight to 11:59pm. For each LAD we identified the set of points on the ERA5 grid contained in the LAD or within 30-km of its boundary. For each weather parameter, we then took the average of all points in the corresponding set, to obtain a LAD level daily average. This strategy provides a consistent rule for assigning measurements of weather data, particularly to LADs that do not contain one of the grid points from the ERA5 data.

\newpage

\textit{\section{Methodology}}
\subsection{Software}
All data processing and analyses were completed in the R language. The \textit{wgrib} software was used to convert the data from ERA5 into a format readable by R. The \textit{CARBayesST} package was used to run MCMC spatio-temporal Bayesian models whilst the packages \textit{ggplot2}, \textit{rgdal}, \textit{leaflet}, \textit{coda}, and \textit{spdep} were used to analyse results.

\subsection{Statistical modelling}
\subsubsection{Bayesian inference}
In Bayesian inference we assume that an observable random variable is generated by a process whose parameters are themselves random variables. Our goal is to study this process by learning about the parameters. In Bayesian inference our learning process starts with our prior belief, before seeing any data, then incorporates observations $y$ of the observable random variable.
Let our observable random variables be $Y$, and $\Theta$ a set of parameters for its distribution $p_Y$. Let $p_{\Theta }(\theta)$ be a joint probability distribution representing our prior belief for the parameters; this may be a non-informative prior if we have no strong belief. The likelihood of an observation $y$ is the conditional distribution $p_{Y}(y|\theta)$. 
Bayes’ Theorem states that we can calculate the joint posterior distribution of the parameters as:
\begin{equation}
    p(\theta|y) \propto p_{Y}(y|\theta)p_{\Theta}(\theta)
\end{equation}
This means that our updated belief after observing $y$, represented by $p(\theta|y)$, can be computed by multiplying our prior belief with the conditional likelihood of the data. It is enough to know the posterior up to a normalising constant since we know the true posterior’s integral over its full support must equal 1.

The idea that parameters are themselves random variables can be extended so that the parameters $\beta_{\theta}$ for the distribution $p_{\Theta}$,  often referred to as hyperparameters, are also random variables. In fact this can be continued indefinitely, however, as more levels are added interpretability can be lost. Because of this at some point hyperparameters are set as point estimates.

\subsubsection{Introducing a lag}
When someone is infected with COVID-19 there is a delay before they become ill known as the virus incubation period. There may then be further delays before they get tested and the result is obtained. As such to assess the impact of weather on COVID-19 transmission risk we introduced a 6-day lag between weather factors and daily cases. In this report days refer to the day cases were recorded, the lag was created by taking the weather from 6 days prior. This effectively reduces the data set to $N = 55$ time periods.

\subsubsection{Model covariates}
In the model we included regression variables for temperature, relative humidity, wind velocity, solar radiation and precipitation. Due to the large positive skew in the wind velocity and precipitation variables we applied a transformation to normalise them. For the wind variable we applied a log transformation and for precipitation a Box-Cox transformation with parameter 0.274, both after replacing zero values with the minimum non-zero value.
One factor not often considered in the existing literature is the difference between weekdays due to administrative processes. Since fewer cases were reported on weekends due to reduced testing capacity \cite{weekendCases} we included weekdays as a factor regression variable to capture these effects. In our analysis, Thursday presents the baseline against which other days are measured. 

\subsubsection{Standardised case rates}
The number of daily cases in a given LAD is highly dependent on the population. To account for this we chose to include expected cases as an offset in the model, allowing us to effectively model the relative risk across all LADs. The model random effects capture latent temporal and spatial trends within the case count data, so the expected count only needs to consider the LAD specific population. Since we consider a relatively small time period we assume LAD population remains constant over time, as it is not practical to obtain daily counts. This means that LAD-specific expected cases remain constant in time.

As such expected cases are defined in equation (2) where $P_k$ is the population of the $k^{\mathrm{th}}$ LAD and $y_{kt}$ its observed daily cases at time $t$.
\begin{equation}
E_{k} = \frac{P_k}{\sum_{k=1}^{K}P_k} \times \frac{1}{N}\sum_{k=1}^{K}\sum_{t=1}^{N}y_{kt}.
\end{equation}

The LAD specific relative risk for time $t$, $\theta_{kt}$, is estimated by:

\begin{equation}
\hat\theta_{kt} = \frac{y_{kt}}{E_{k}}.
\end{equation}

\subsubsection{Bayesian spatio-temporal model with Adaptive Prior}
As the number of cases $Y_{kt}$ in LAD $k$ at time $t$ can be modelled as count data we use a Bayesian hierarchical Poisson GLM for the daily cases likelihood function. It is common for count data to be generated from several underlying Poisson processes, leading to `over-dispersion' where the variance in the data is much higher than the mean, so that the data is not well modelled by a single Poisson distribution. However, in our Poisson GLM the Poisson parameter is the product of expected cases $E_{k}$ and relative risk $\theta_{kt}$ for the $k^{\mathrm{th}}$ local authority at time $t$. The relative risk is dependent on a linear predictor containing covariates $\mathbf{x}_{kt}$ and spatio-temporal random effects $\phi_{kt}$. As the relative risks vary spatially and temporally through the covariates and spatio-temporal effects our model fits a different Poisson distribution to each LAD-time pair. This allows us to capture the underlying system behaviour and its corresponding over-dispersion.

In the linear predictor $\mathbf{x}_{kt}$ denotes model covariates with corresponding regression parameters $\beta$ given a non-informative prior. 
The spatio-temporal random effects $\phi_{kt}$ act as a latent variable for factors affecting the relative risk $\theta_{kt}$ that are not given as model inputs. For instance, these will adjust for case spread due to the number of cases in LAD $k$, and neighbouring LADs, at time $t-1$. As such these random effects are correlated both spatially and temporally. Different structures can be given to these random effects, all of which use a neighbourhood matrix $W$ to identify geographically correlated data points \cite{lee2022spatio}, but have other differing underlying assumptions on the system behaviour.

In this study we use an adaptive prior which assumes the auto-correlated random effects follow a normal distribution whose mean of $\Phi_t$, the collection of random effects at time $t$, depends only on the random effects at the previous time period, with the first time period having a zero mean. The adaptive prior allows us to replace non-zero elements of $W$ with random variables $\mathbf{w}^+$. Through a precision matrix $\mathbf{Q}$ these control localised spatial correlation between random effects. As they are random variables they allow different LAD pairs to have different correlation strengths. For example, correlation may be greater between neighbouring inner-city LADs than between neighbouring rural LADs.

The precision matrix also depends on the parameter $\rho_S \in [0,1]$ which controls the underlying strength of spatial variance whilst the overall spatial variance also depends on the variance parameter $\tau^2$. In the $\tau^2$ Inverse-Gamma prior the default values of $a,b$ are respectively 1 and 0.01. Details of the Inverse-Gamma distribution are presented in the Appendix. The strength of temporal dependence in random effects is controlled by $\rho_T$ which has support $[0,1]$.  

Given $P$ covariates the model is then defined as:
\begin{equation}
\begin{split}
    Y_{kt} &\sim \mathrm{Poisson}(E_{k}\theta_{kt}) \ \ \ \ \ \ \ \  k = 1,\cdots,K, \  \ t = 1,\cdots,N \\
    \log(E_k\theta_{kt}) &= \log(E_k) + \beta \mathbf{x}_{kt} + \phi_{kt} \\
    \mathbf{x}_{kt} &= (x_{kt1},\cdots,x_{ktP})^T \\
    \beta &= (\beta_1,\cdots,\beta_P)\\
    \beta_p & \sim N(0,10^5) \ \ \ \ \ \ \ \ \ \ \ \ \ \ \ \ \  p = 1,\cdots,P\\
    \Phi_t|\Phi_{t-1} & \sim \mathbf{N}(\rho_T\Phi_{t-1},\tau^2\mathbf{Q}(W,\rho_S)^{-1}) \ \ \ \ \ \ t = 2,\cdots,N \\
    \Phi_{1} & \sim \mathbf{N}(\textbf{0},\tau^2\mathbf{Q}(W,\rho_S)^{-1}) \\
    \tau^2 & \sim \mathrm{InverseGamma}(a,b) \\
    \rho_S, \rho_T & \sim \mathrm{Uniform}(0,1)\\
     \mathbf{Q}(W,\rho_S) &= \rho_S(\mathrm{diag}(W\mathbf{1}) - W) + (1 - \rho_S)I_K.
\end{split}
\end{equation}

To model $\mathbf{w^+}$ as a collection of random variables they're first mapped to the real line via the following transformation:
\begin{equation}
    \mathbf{v^+} = \log(\mathbf{w^+} / (1 - \mathbf{w^+})).
\end{equation}
A multi-variate Gaussian prior with fixed mean $\mu$ and inverse-gamma variance $\tau_{\omega}^2$ is used to model the transformed elements so the joint density $f$ is:
\begin{equation}
\begin{split}
    f(\mathbf{v^+}|\mu,\tau_{\omega}^2) & \propto \exp\left(-\frac{1}{2\tau_{\omega}^2}\sum_{v_{ij}^+\in\mathbf{v^+}}(v_{ij}^+ - \mu)^2\right)\\
    \tau_{\omega}^2& \sim \mathrm{InverseGamma}(c,d)
\end{split}
\end{equation}

Following the work of Rushworth in \cite{rushworth2017adaptive} the prior choice of the value $\mu$ influences the type of posteriors found for elements of in the solutions found $\mathbf{w^+}$. Values greater than 0 generally prefer posterior elements close to 1, indicative of auto-correlation. In particular, the default value of 15 assigns high probability for elements being close to 0 or 1, corresponding to neighbouring LADs displaying correlation or not. Since the case counts display spatial auto-correlation we proceed with the default value. The default hyperparamater values $c, d$ in the $\tau_{\omega}^2$ prior are respectively are 1 and 0.01.

The variance parameter $\tau_{\omega}^2$ controls the amount of variation in the $\mathbf{v^+}$ and hence $\mathbf{w^+}$. Small values of $\tau_{\omega}^2$ make the density value small almost everywhere, except when $v_{ij}^+$ is close to $\mu$. Therefore small values correspond to globally strong auto-correlation whilst large values allow auto-correlation strength to vary between neighbours.

\subsubsection{Neighbourhood matrix}
We defined $W$ using a binary approach identifying neighbouring LADs, an established approach in the use of Bayesian spatio-temporal models, which introduces spatial correlation in the random effects between neighbouring LADs. If $LA_i$ denotes the $i^{\mathrm{th}}$ local authority the neighbourhood matrix is such that:

\begin{equation}
w_{ij} := W[i,j] = {\begin{cases}{1}&\mathrm{if} \ LA_{i} \ \& \ LA_{j} \ \mathrm{are} \ \mathrm{neighbours}\\
0 & \mathrm{otherwise}\end{cases}}
\end{equation}

Since the model requires a contiguous geography we defined the neighbours of the Isle of Wight to be those in England with ferry routes to the island, namely Portsmouth, Southampton and the New Forest. 

\subsubsection{Second neighbour model}
To investigate the effect of the neighbourhood matrix specification and better understand the spatio-temporal effects we ran a second model ($M_1$) using the same structure as the original model ($M_0$) but with a different neighbourhood matrix $W'$. This was based on the original neighbourhood matrix but with an adaptation made for England's five largest metropolitan areas, Greater London, Tyneside, Merseyside (excluding The Wirral), Greater Manchester and the West Midlands. Within each of these regions we identified two non-neighbouring LADs as neighbours if they shared a common immediate neighbour also within that region.

\subsubsection{Model implementation}
The adaptive model was implemented using version 3.3 of the \textit{CARBayesST} package in R. Since MCMC methods are computationally intensive we ran four MCMC chains in parallel, each with a burn-in of 200,000 samples and a further 500,000 samples, thinned by 200 to reduce sampling auto-correlation. This yielded a total of 10,000 samples (2,500 per chain) for inference. All chains used the default fixed hyperparameters whilst regression parameters were given the default normal prior distribution of mean 0 and variance $10^5$. The regression parameters were sampled using the simple random walk algorithm method within the \textit{CARBayesST} package. 

We checked the posterior distributions for convergence of individual chains using the Geweke diagnostic and for convergence between MCMC chains using the Gelman-Rubin diagnostic. Our inference is based on the model's 95\% highest posterior density intervals and posterior means. 

\subsubsection{Spatio-temporal diagnostics}
To validate the performance of our models we examined the residuals for spatio-temporal auto-correlation. We tested for spatial auto-correlation using Moran's I statistic, with the binary neighbourhood matrix. This test was applied for each time period individually and a Bonferroni correction was used for inference. Temporal auto-correlation was tested for using the Ljung-Box test for time series auto-correlation for each LAD. Again, a Bonferroni correction was used for inference.

\newpage

\section{Results}

After adjusting the data to account for lags between weather data and daily cases we were left with 55 time periods and 312 LADs, giving 17,160 total observations. Key summary statistics on the data are presented in Table 1 below: 

\begin{table}[ht]
\caption{\label{tab:SummaryStats}Summary statistics for cases and meteorological variables.}
\centering
\begin{tabular}{|c||c|c|c|c|c|}\hline
\textbf{Variable} & \textbf{Min.} & \textbf{Median} & \textbf{Mean} & \textbf{Max.} & \textbf{S.D.} \\\hline
Cases & 0.0 & 6.0 & 9.5 & 131 & 10.6 \\\hline
Temperature & 4.0 & 9.0 & 9.3 & 17.0 & 2.2 \\\hline
Relative humidity & 58.3 & 77.1 & 76.5 & 93.4 & 5.9 \\\hline
Wind velocity & 1.6 & 5.2 & 5.5 & 14.2 & 2.0 \\ \hline
SSRD & 0.6 & 1.5 & 1.5 & 2.4 & 0.3 \\\hline
Precipitation & 0.0 & $2.4 \times 10^{-2}$ & $4.7 \times 10^{-2}$ & $5.6 \times 10^{-1}$ & $6.4 \times 10^{-2}$\\\hline
Wind velocity (log) & 0.5 & 1.7 & 1.6 & 2.7 & 0.4 \\ \hline
Precipitation (Box-cox parameter: 0.274) & -3.65 & -2.26 & -2.28 & -0.53 & 0.6 \\\hline
\end{tabular}   
\end{table}

An initial visualisation of the data in Figure 1 shows fewer cases are regularly reported on Saturday and Sunday of each week, whilst all other days appear to be relatively consistent in reported cases. Geographically, Figure 2 displays the relative risk across the study period by LAD on a $\log_2$ scale, so that an increase by one unit corresponds to a doubling in the relative risk. This displays clear clustering in the relative risk with high risk areas including Kent, Manchester and Tyneside, whilst North Lincolnshire and the Southwest of England are areas of low risk. 
\begin{figure}[ht]
\textbf{Mean LAD cases by weekday during the study period (Apr - May)}
\centering
\includegraphics[height=10.0cm]{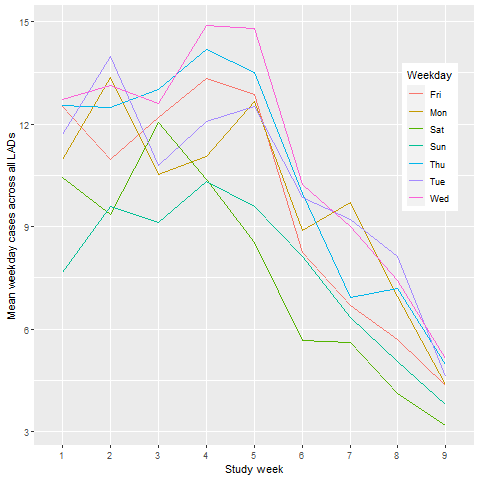}
\caption{Mean weekday cases show a drop in case counts on the weekend}
\end{figure}

\begin{figure}[ht]
\centering
\textbf{Mean relative risk for each LAD across the study period}
\includegraphics[height=10.0cm]{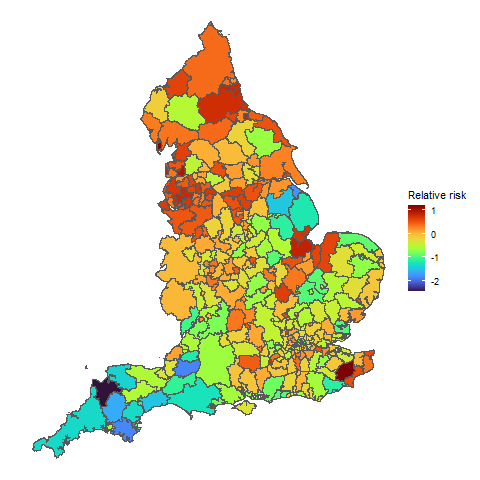}
\caption{Mean relative risk between LADs across the study period displays geographical correlation}
\end{figure}

\subsection{Model inference}

\begin{figure}[h!]
\centering
\textbf{Posterior distributions for weather and intercept parameters with 95\% credible intervals}
\includegraphics[height=10.0cm]{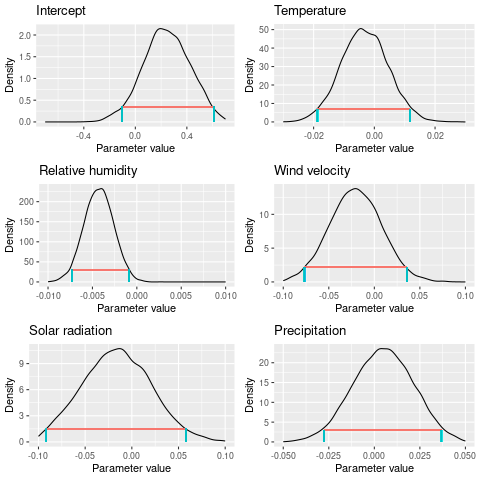}
\caption{Only humidity is associated with daily cases at a 95\% credible level}
\end{figure}

\subsubsection{Weather analysis}
The posterior distributions in Figure 3 for each of temperature, wind velocity, solar radiation and precipitation show no strong relationship between the covariate and the number of observed cases. Whilst the 95\% HPD for relative humidity lies below 0 the size of the parameter in this interval suggests the effect is marginal, with a 1\% rise in humidity corresponding to a fall in the relative risk between 0.1\% - 0.3\%.

\subsubsection{Spatial analysis}
Qualitative analysis of the model residuals suggests it's largely captured spatial trends in the data as the mean residuals for each LAD have a random pattern, seen in Figure 4. Quantitatively, using Moran's I statistic at a 95\% significance level with a Bonferroni correction, no time period showed spatial auto-correlation.

The posterior of $\rho_S$ in Figure 5, controlling underlying the strength of spatial auto-correlation, is close to 1. This allows for strong spatial auto-correlation in the random effects $\phi_{kt}$. However, to understand the spatial structure the $\mathbf{w}^+$ elements must also be considered. The variation amongst $\mathbf{w}^+$ elements, controlled by $\tau_{\omega}^2$, is large taking values in the range of 150 - 220. This large variation allows for both strong and weak correlations in the neighbourhood matrix posteriors. The posteriors for both $\rho_S$ and $\tau_{\omega}^2$ remain similar in the second model.

In both versions of the model the majority of $\mathbf{w}^+$ elements have a posterior density towards 1, a few elements have most of their density towards 0. This demonstrates the model's identifying an overall geographical correlation in daily cases, but finding there are a small number of neighbouring LADs that did not influence each other.

Whilst on a national scale the collection of spatial correlations display similar trends between the two models, there are differences at a localised level. In $M_1$ there are both examples of strong second neighbour correlations and differences in immediate neighbour correlations from $M_0$.  

An example of this is in Hackney, which Figure 6 shows was strongly correlated with all 9 of its neighbouring LADs in $M_0$. This gives credence to our assumption of high correlation in urban areas. However, whilst $M_1$ finds Hackney to be closely correlated with Camden it gives a different view on other neighbours as shown in Figure 7. For instance, Hackney is mostly uncorrelated with Westminster. In contrast, Figure 8 shows Hackney is strongly correlated with its Eastern second neighbours Barking \& Dagenham and Redbridge, whilst the correlations are less strong with southern second neighbours, in particular Bromley.

Figure 7 also highlights that $\mathbf{w}^+$ elements can have bimodal posteriors, such as the correlation between Hackney and Islington. This style of posteriors allows these LADs to share an overall trend with some elements of independent behaviour. Figure 9 displays samples from the random effect posteriors for Hackney and Islington. Whilst both LADs have an overall decreasing trend in terms of relative risk, Islington has a relatively stationary random effect in days 5-12 and 28-38. Contrastingly, Hackney's random effect is stationary for a much smaller time, generally decreasing in these periods. 

Whilst the localised correlation structures differ between $M_0$ and $M_1$, we found the overall regression results were insensitive to the change in neighbourhood matrix. In both models humidity displayed a small effect at a 95\% confidence level, with other factors showing no evidence of an association with daily cases.

\begin{figure}[ht]
\centering
\textbf{Mean LAD residuals across the study period}
\includegraphics[height=10.0cm]{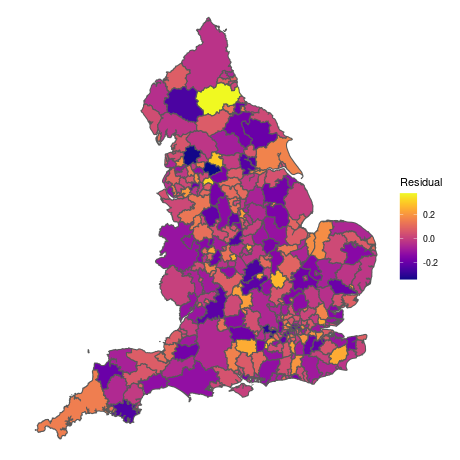}
\caption{Mean residuals by LAD shows the model has captured geographical trends in the data}
\end{figure}

\begin{figure}[h!]
\centering
\textbf{Posterior distributions of underlying correlations and variance hyperparameters}
\includegraphics[height=10.0cm]{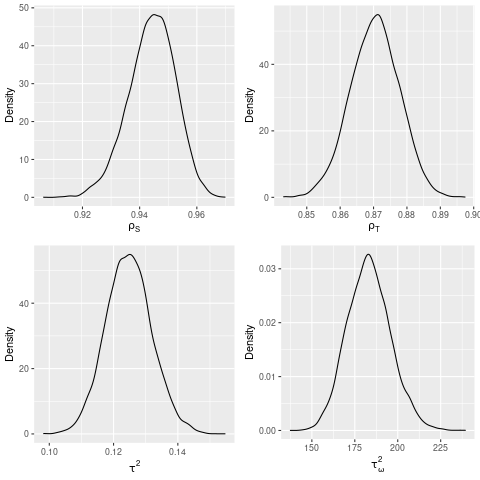}
\caption{Hyperparameter posteriors show strong spatio-temporal auto-correlation}
\end{figure}

\begin{figure}[ht]
\centering
\textbf{Posterior correlations between Hackney and immediate neighbours in $M_0$}
\includegraphics[height=10.0cm]{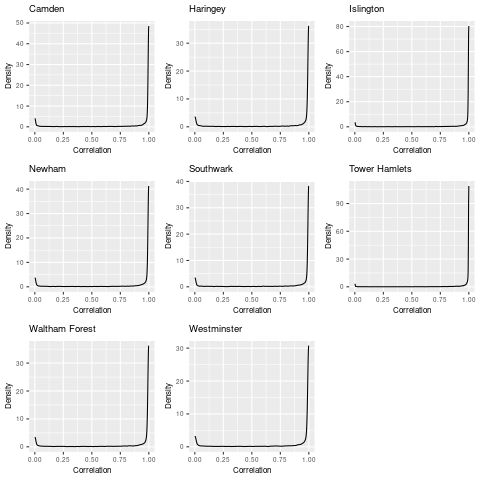}
\caption{In $M_0$ Hackney is strongly correlated to all of its immediate neighbours}
\end{figure}

\begin{figure}[h!]
\centering
\textbf{Posterior correlations between Hackney and immediate neighbours in $M_1$}
\includegraphics[height=10.0cm]{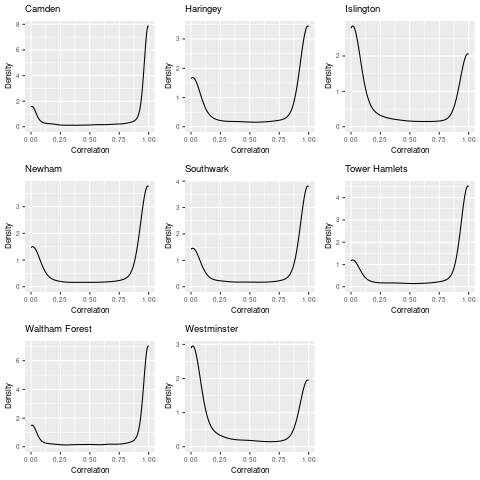}
\caption{Model $M_1$ finds Hackney to be less strongly correlated with its immediate neighbours}
\end{figure}

\begin{figure}[ht]
\centering
\textbf{Posterior correlations between Hackney and second neighbours in $M_1$}
\includegraphics[height=10.0cm]{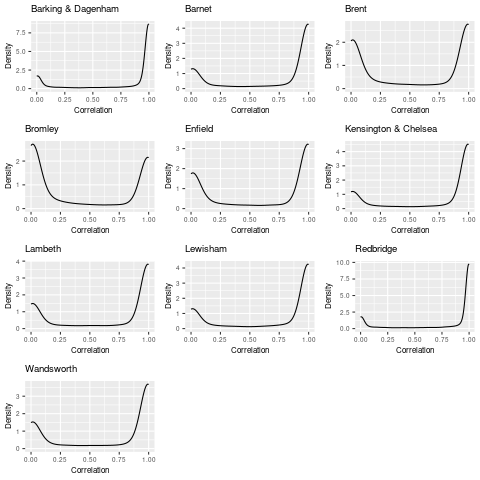}
\caption{In $M_1$ Hackney is correlated to several second order neighbours}
\end{figure}

\begin{figure}[h!]
\centering
\textbf{Hackney and Islington random effect samples in $M_1$}
\includegraphics[height=10.0cm]{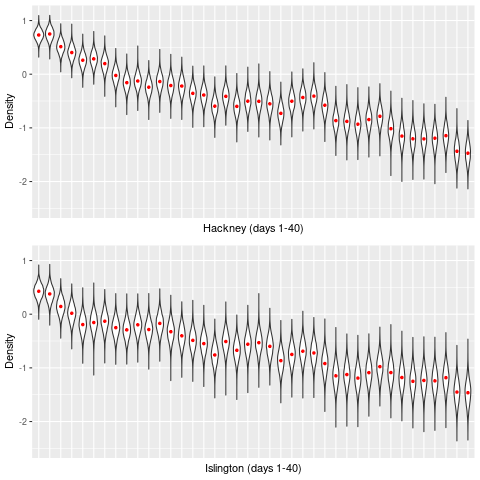}
\caption{In $M_1$ the Hackney \& Islington random effects share a common trend with localised differences}
\end{figure}

\newpage
\subsubsection{Temporal analysis}
Overall the model captures the temporal trend in the data. Figure 10 shows the mean daily residuals across LADs fluctuating around 0 with no discernible difference between weekdays. The temporal trend within individual LADs is also captured, with the Ljung-Box test indicating just 4.8\% of LADs showing temporal auto-correlation in their residuals at a 95\% significance level.

The weekday posteriors in Figure 11 highlight the trend of fewer cases being reported on weekends. with both Saturday and Sunday having large negative coefficients. Their 95\% posteriors are in the ranges  (-0.437, -0.272) and (-0.448, -0.263) respectively. This corresponds to, relative to Thursday, 24-35\% fewer cases reported on Saturdays with 23-36\% fewer cases reported on Sundays. Friday also shows fewer cases reported to a lesser extent, with between 3-15\% fewer cases. There is no evidence of consistent case over-reporting on any particular day, with all other weekday 95\%  highest posterior densities containing 0. 

The posterior for $\rho_T$ in Figure 5 is clustered around 0.85 - 0.89. This introduces high correlation in the random effects $\phi_{kt}$. However, the temporal variance between the random effects, controlled by $\tau^2$, is relatively small clustered around 0.11 - 0.14. The effect of these hyperparameters is seen in the Hackney and Islington random effects, where the random effects for each LAD change smoothly and gradually over the 40-day period.

\begin{figure}[ht]
\centering
  \textbf{Mean daily residuals across LADs with boxplot of all LAD residuals by weekday}
\includegraphics[height=10.0cm]{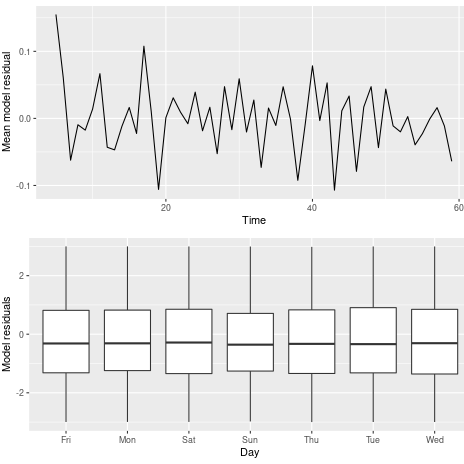}
\caption{Mean daily residuals show a random pattern with no clear distinction between weekdays}
\end{figure}

\begin{figure}[h!]
  \centering
  \textbf{Posterior distributions of weekday parameters with 95\% credible intervals}
  \includegraphics[height=10.0cm]{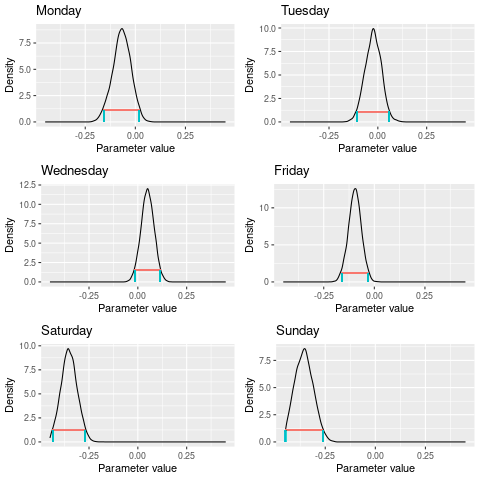}
  \caption{Weekday posteriors demonstrate fewer cases are reported at weekends.}
  \label{fig:label-name}
\end{figure}

\newpage
\section{Discussion}
In this report we aimed to identify relationships between meteorological factors and COVID-19 transmission using Bayesian spatio-temporal techniques. We've shown this type of model can capture the auto-correlation present in observed cases of a human-transmissible virus. Our results provide evidence for relative humidity being negatively associated with COVID-19 transmission but no evidence that other factors are associated with the virus's spread. We believe that weather conditions are unlikely to critically affect COVID-19 transmission.

The model posterior provides evidence for a negative relationship between relative humidity and COVID-19 spread, something that is disputed amongst previous studies. For instance, in Germany \cite{ganegoda2021interrelationship} claimed a positive correlation whilst in the US \cite{chien2021lagged} proposed a negative relationship. More widely, \cite{Ai2022} found both low and high ranges of humidity could be associated with increased COVID-19 prevalence.

Our findings provide no evidence that other meteorological are associated with COVID-19 transmission. For temperature this agrees with the findings of \cite{briz2020spatio} in Spain, despite their study's use of a non-monotonic relationship between temperature and observed cases, allowing for a possible `peak transmission' range to be identified.

A possible reason for the lack of meteorological effects is that initially the virus's spread may have been driven almost entirely by human interaction due to the low immunity in the population. Whilst spatio-temporal effects are used to account for these factors we appeal to George Box's famous quote that ``All models are wrong, but some are useful''. We interpret this as believing the spatio-temporal effects help to understand the spatial and temporal structure of the relative risk. However, the `true' effect on the spread of COVID-19 may be so sensitive to slight changes not captured in the model. For instance, the complexities of how society adjusted to lockdowns, the increase in testing capacity during the study period as well as changes in case reporting whilst the infrastructure was developed could all impact the observed case numbers. 

The main feature of the adaptive spatio-temporal effect structure is that it allows the model to account for different strengths of spatial auto-correlation across the full geography. As shown in \cite{rushworth2017adaptive} and this report the model design encourages values close to 1, representing correlation, or 0 representing independence. However, by limiting spatial auto-correlation through $\rho_S$ and allowing for spatial variation through $\tau_{\omega}^2$ the model remains flexible enough to capture localised trends even when two neighbours are highly correlated.

We found that a change in the neighbourhood matrix allowed us to place different assumptions on the underlying behaviour. In particular, the use of a more complex neighbourhood matrix could enable researchers to discover more complex relationships between LADs. However, since the overall regression results are insensitive to our change we believe it's not crucial to find a `perfect' neighbourhood matrix to analyse a problem. This allows researchers to make simplifying assumptions on geographical dependencies and design their neighbourhood matrix appropriately on that basis. This could be advantageous when either finding a `perfect' neighbourhood matrix is a complex problem, or using one would put strain on computational resources. 

One limitation of our findings is that we have not compared our results with different models, such as a spatial moving average model. One interesting approach would have been to use an Integrated Nested Laplace Approximation (INLA) model. The key advantage of INLA over MCMC methods is its low computational cost, since it focuses on approximating the marginal posteriors. Whilst INLA does not provide the same theoretical guarantees as MCMC, in practise it has proven to be an effective tool. However, one limitation of INLA is that it should only be used with a small number of hyperparameters outside of the linear predictor, otherwise the computational cost savings are lost \cite{rue2016bayesian}. In particular, we could not natively use INLA to model the underlying spatial correlations between LADs, as we did using the adaptive prior in \textit{CARBayesST}. We believe this capability to identify a more complex spatial structure provides greater insight into COVID-19 incidence in this scenario.

Overall this report has found that there is no evidence of an association between most meteorological factors and COVID-19 spread. In particular, this report shows that COVID-19 transmission may be impacted by humidity, but is unlikely to be critically limited by it. As such public policies designed to address COVID-19 should not rely on seasonality as part of their strategy.

\newpage
\section{Conclusion}
In this study, we accounted for the nature of the virus in two ways. By introducing a lag between the variables and observed cases we adjusted for the delay between transmission and reporting cases due to an incubation period. Other authors such as \cite{ganegoda2021interrelationship} have shown how this can be extended to include multiple weighted lags. We also controlled for correlation in virus transmission through the use of spatio-temporal effects. We've shown how our model has captured both temporal and spatial trends in daily cases, allowing us to make robust conclusions from the model outputs.

This study focused on isolating the main effects of individual meteorological factors on COVID-19 spread. To do so we looked at a narrow time frame from $31^{\mathrm{st}}$ March 2020 until $30^{\mathrm{th}}$ May 2020, covering only the initial phase of the pandemic. This has several important implications regarding the generalisability of the results. Firstly, results are derived from limited ranges for the meteorological variables, in particular temperature, solar radiation and humidity. As such the results are only applicable in countries with similar climates and may not apply to England during other seasons. It would be prudent for future studies to include data from a full year to better assess the effect of meteorological variables. 

A more complex version of our model could introduce a spatial, temporal or spatio-temporal regression parameters for each meteorological variable to identify whether any weather conditions have spatial, temporal or temporal-spatial associations with the spread of COVID-19. However, we believe any associations found would both be difficult to interpret and less likely to generalise outside of the specifics of our study. As such we chose to focus solely on the average effect of each factor. 

The introduction and subsequent successful uptake of COVID-19 vaccines, allowing the lifting of lockdown restrictions, has dramatically altered where and when the virus can spread. This is likely to have had an impact on how important meteorological factors are associated with COVID-19 transmission. In England all legal social restrictions were lifted by the end of February 2022, but it may be of interest to perform similar analyses with 12 months of data where no restrictions were in place. Whilst vaccinations continue during this period the vast majority of adults had received the recommended 2 doses by this point.

Future research could also consider different design principles for the neighbourhood matrix. In this study the neighbourhood matrix was based solely on neighbouring local authorities, the most common approach in the literature. Another simplistic option would be to have a weight between all pairs of local authorities that is inversely related to distance, justified by assuming the further apart two local authorities are the less related they are likely to be. A more complex approach would be to estimate the amount of human traffic between neighbouring local authorities and define weights based on this. Dependent on the alternative approach this may require a different prior structure for the spatio-temporal effects as the adaptive model requires a binary neighbourhood matrix.

\newpage

\section{Appendix}
\subsection{Inverse-Gamma Distribution}
A variable $X$ follows an Inverse-Gamma($a$, $b$) distribution if it's support is the positive real line and it has a probability density function 
$$ f(x, a, b) = \frac{b^a\mathrm{e}^{-b/x}}{x^{a+1}\Gamma(a)}$$
where $\Gamma(\cdot)$ is the Gamma function. Both the shape parameter $a$ and scale parameter $b$ must be positive.

\subsection{Meteorological variables}
\begin{itemize}
    \item \textbf{Temperature} is given at 2-metres above the Earth's surface in Kelvin. In the study we converted this to Celsius.
    
    \item \textbf{Humidity} is measured by relative humidity.
   
   \item \textbf{Wind velocity} is given at a height of 10-metres above the Earth's surface. 
   
   \item \textbf{Solar radiation} is defined in terms of surface solar radiation downwards, measured in energy per unit area. Due to the large values, this variable was scaled down by a factor of $10^6$.
   
   \item \textbf{Precipitation} is measured as the cumulative depth of daily precipitation spread over the geographical region in metres. Due to the small units we scaled this by 1,000, effectively measuring depth in millimetres.
\end{itemize}

\subsection{A word on notation}
In section 3.2 $Y_{kt}$ is a random variable for the number of cases in the $k^\mathrm{th}$ LAD at time $t$, whereas $y_{kt}$ refers to the observed number of cases. Similarly, $\theta_{kt}$ is a random variable for the corresponding relative risk whilst $\hat{\theta}_{kt}$ is a point estimate of the relative risk. 

\subsection{Convergence diagnostics}
Below are the convergence diagnostics for model $M_0$. The Gelman-Rubin diagnostic is denoted G-R, and $\mathrm{G}_i$ is the Geweke diagnostic for the $i^\mathrm{th}$ chain. 
\begin{table}[h]
\caption{\label{tab:RegParams} Convergence of models for regression parameters.}
\centering
\begin{tabular}{|c||c|c|c|c|c|}\hline
\multirow{2}{55pt}{\textbf{Feature}} &  \multicolumn{5}{|c|}{\textbf{Diagnostic}}\\
& G-R & $G_1$ & $G_2$  & $G_3$ & $G_4$\\\hline
Intercept & 1.01 & 1.4 & -1.6 & -1.4 & 0.1
\\\hline
Monday & 1.01 & -2.2 & 1.1 & 0.9 & -1.8
\\\hline
Tuesday & 1.01 & -2.6 & 1.6 & 1.1 & -2.1
\\\hline
Wednesday & 1.00 & -2.2 & 2.2 & 1.2 & -1.1
\\\hline
Friday & 1.01 & -1.0 & 0.8 & 1.5 & -0.7
\\\hline
Saturday & 1.01 & -0.8 & 1.2 & 1.4 & -0.5
\\ \hline
Sunday & 1.01 & -1.8 & 1.2 & 1.3 & -0.6
\\\hline
Temperature & 1.00 & 0.1 & 1.5 & 1.4 & -0.3
\\\hline
Relative humidity & 1.00 & -0.6 & 0.9 & -0.5 & 0.9
\\\hline
Wind velocity & 1.00 & -0.4 & -0.3 & 1.7 & -0.6
\\ \hline
SSRD & 1.00 & -0.2 & -0.1 & -0.1 & -0.5
\\\hline
Precipitation & 1.00 & 0.0 & -0.4 & -0.4 & -1.4 \\\hline
$\rho_S$ & 1.00 & 0.5 & -0.4 & 0.6 & -0.3
\\\hline
$\rho_T$ & 1.00 & 1.4 & 0.2 & 0.2 & -0.4
\\ \hline
$\tau^2$ & 1.00 & -1.1 & -0.3 & -0.7 & -0.1
\\\hline
$\tau_{\omega}^2$ & 1.00 & 0.9 & -0.2 & 0.8 & -0.4
\\\hline

\end{tabular}   
\end{table}

\newpage
\subsection{Trace Plots}
Reading left to right, up to down, the trace plots are the posterior samples for Sunday, Friday, Saturday, Wednesday, Monday and Tuesday. 
\begin{figure}[h!]
\centering
\textbf{MCMC trace plots for weekday parameters}
\includegraphics[height=11.0cm]{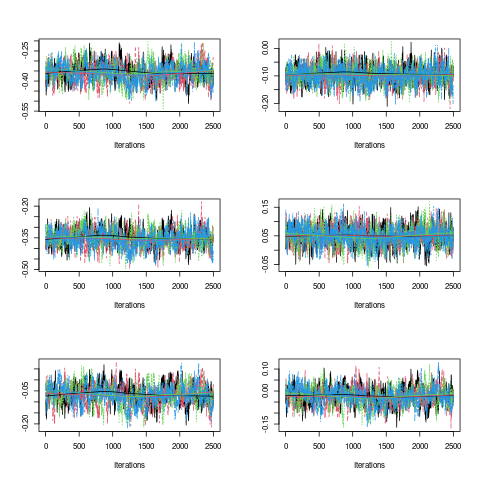}
\caption{Trace plots demonstrate weekday parameters may not have converged}
\end{figure}

\newpage
Reading left to right, up to down, the trace plots are the posterior samples for the intercept, temperature, relative humidity, wind velocity, solar radiation and precipitation.

\begin{figure}[h!]
\centering
\textbf{MCMC trace plots for intercept and meteorological parameters}
\includegraphics[height=11.0cm]{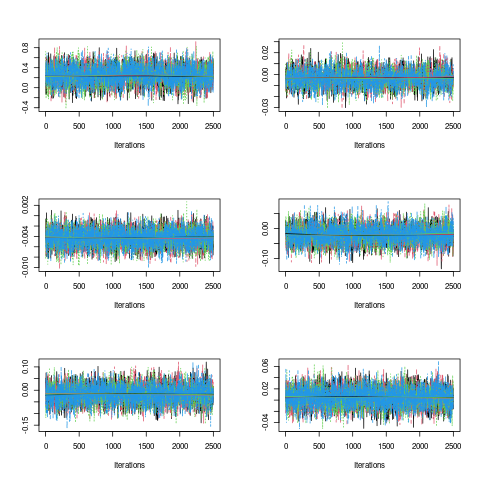}
\caption{Trace plots provide strong evidence weather parameters have converged}
\end{figure}

\newpage

\section*{Acknowledgements}
The authors thank David Whitney for his help in pre-processing the raw data.

\section*{Funding}
Nothing to declare.

\section*{Abbreviations}
\begin{itemize}
    \item ECMWF: European Centre for Medium-Range Weather Forecasts
    \item LAD: Local Authority District
    \item LTLA: Lower Tier Local Authority
    \item MCMC: Markov chain Monte Carlo
    \item ONS: Office for National Statistics
    \item S.D.: Standard Deviation
\end{itemize}

\section*{Availability of data and materials}
The datasets supporting the conclusions of this article are included within the article and its additional files.

\section*{Ethics approval and consent to participate}
Not applicable.

\section*{Competing interests}
The authors declare that they have no competing interests.

\section*{Consent for publication}
Not applicable.

\section*{Authors' contributions}
T. Jendoubi conceived the research and cleaned the data. The research was guided by T. Jendoubi and B Leurent; J. Mullineaux analysed the data and wrote the manuscript. T. Jendoubi and B. Leurent reviewed and edited the manuscript. All authors have read and approved the manuscript.

\section*{Authors' information}
Authors and Affiliations

\textbf{Department of Statistical Science, University College London, Gower Street, London, WC1E 6BT, United Kingdom}

Jamie Mullineaux, Baptiste Leurent and Takoua Jendoubi

\bibliographystyle{reference-style}
\bibliography{report}

\section*{Additional Files}
\subsection*{Additional file 1 --- W.rds}
An R object that can be loaded using the \textit{readRDS} command. The loaded object is a binary matrix used to define neighbouring LADs in the original model. 

\subsection*{Additional file 2 --- new\_W.rds}
An R object that can be loaded using the \textit{readRDS} command. The loaded object is a binary matrix used to define neighbouring LADs in the extended model. 

\subsection*{Additional file 3 --- LoadData.Rmd}
R-Markdown file containing code to load the cleaned data sources, prepare the data for analysis and load required libraries.
    
\subsection*{Additional file 4 --- ExploratoryDataAnalysis.Rmd}
R-Markdown file containing code to carry out an exploratory analysis of the prepared data.

\subsection*{Additional file 5 --- ParallelModel.Rmd}
R-Markdown file containing code to set-up, run both and save the main model using parallelisation.

\subsection*{Additional file 6 --- SensitivityModels.Rmd}
R-Markdown file containing code to set-up, run and save the extended model incorporating second-order neighbours. Makes use of parallelisation.
    
\subsection*{Additional file 7 --- ParallelAnalysis.Rmd}
R-Markdown file containing code to analyse both models. Includes model diagnostics, analysis of reisduals and investigations into both model parameters and hyperparameters.
    
\subsection*{Additional file 8 ---RandomEffectAnalysis.Rmd}
R-Markdown file containing code to conduct analysis into the random effects in both models.

\subsection*{Additional file 9 --- CleanWeatherData.rds}
An R object that can be loaded using the \textit{readRDS} command. Contains daily measurements of for each LAD. LADs are identified by name and LAD code.

\subsection*{Additional file 10 --- CasePopulationData.csv}
Contains daily case counts for each LAD along with the LAD population from the ONS estimates. LADs are identified by LAD code and name.
    
\subsection*{Additional file 11 --- LAD\_(Dec\_2020)\_UK\_BFC.zip}
A zipped-folder containing the shapefiles required to create the maps of English LADs.

\end{document}